\documentclass[twocolumn,showpacs,amsmath,amssymb,prl,superscriptaddress]{revtex4-1}
\usepackage{graphicx}
\usepackage{dcolumn}
\usepackage{bm}

\newcommand{\gras}[1]{\boldsymbol{#1}}

\begin{document}

\title{Microscopic modeling of mass and charge distributions in the spontaneous fission of $^{240}$Pu}

\author{Jhilam Sadhukhan}
\affiliation{Physics Group, Variable Energy Cyclotron Centre, 1/AF Bidhan Nagar, Kolkata 700064, India}
\affiliation{Department of Physics and Astronomy and FRIB Laboratory,
Michigan State University, East Lansing, Michigan 48824, USA}

\author{Witold Nazarewicz}
\affiliation{Department of Physics and Astronomy and NSCL/FRIB Laboratory,
Michigan State University, East Lansing, Michigan 48824, USA}
\affiliation{Institute of Theoretical Physics, Faculty of Physics,
University of Warsaw, 02-093 Warsaw, Poland}

\author{Nicolas Schunck}
\affiliation{Nuclear and Chemical Science Division, Lawrence Livermore National Laboratory, Livermore, California 94551, USA}

\date{\today}

\begin{abstract}
In this letter, we outline a methodology to calculate microscopically mass and charge 
distributions of spontaneous fission yields. We combine the  multi-dimensional minimization of collective action for fission 
with stochastic Langevin dynamics to track the relevant fission paths from the 
ground-state configuration up to scission. The nuclear potential energy and 
collective inertia governing the tunneling motion are obtained with nuclear
density functional theory in the collective space of shape deformations and pairing. We obtain a 
quantitative agreement with experimental data and find that both the charge and mass 
distributions in the spontaneous fission of $^{240}$Pu are sensitive both to the dissipation in collective motion and to adiabatic characteristics. 
\end{abstract}



\maketitle


{\it Introduction} --- Spontaneous fission (SF) is a fundamental radioactive decay 
of very heavy atomic nuclei \cite{(Fle40)}. In basic science, it is a major driver determining the 
stability of the heaviest and superheavy elements \cite{(Nil69),seaborg1990,Oganessian07}. Information on SF rates and 
fission fragment distributions are key ingredients of reaction network 
calculations aiming at simulating the formation of elements in the universe 
through nucleosynthesis processes \cite{panov2005,beun2008,Erl12,(Gor13)}. In the context of 
the nuclear data program, SF data are crucial
for calibration of nuclear material counting techniques relevant to international 
safeguards \cite{nichols2008,Murray2014}. Since the discovery of SF in 1940, considerable 
experimental effort has been devoted to obtaining precise data on SF 
observables such as fission half-lives, fission yield properties
(charge, mass, excitation energy, etc.), and  gamma and particle spectra. However, many nuclei relevant to nuclear 
astrophysics are very short-lived and out of experimental reach. Moreover, measurements in actinide nuclei for nuclear technology applications
can pose safety issues. Theory is, therefore, indispensable to fill in the 
gaps in nuclear data libraries.

Modeling SF represents a daunting theoretical challenge. Fission is an
extreme manifestation of quantum tunneling in a many-body system of strongly interacting particles.
Since fission is believed to be a fairly slow process 
driven  by a few collective degrees of freedom, the most advanced theoretical 
efforts today  are often based on the adiabatic approximation  as implemented in nuclear 
density functional theory (DFT). This approach has proven  successful  in describing SF half-lives \cite{(Sta13),(Giu14),(Rob13),(Cha05)}, but has rarely been considered for the distribution of charge, mass, and kinetic energy of the  SF yields \cite{(Bon97)}.
Even semi-phenomenological models of fission dynamics have been mostly focused on neutron- and gamma-induced fission, and electron-capture delayed fission, but not SF ~\cite{(Ich09),(Ran11),(Ran13),(Mol15),(Ada13),(Lem15),(Sca15)}. Today, empirical scission-point models are the only tools available to calculate SF fragment distributions \cite{(Gor13),wilkins1976}.

Within the DFT picture, the evolution of the nuclear system in SF
can be viewed as a dynamical two-step process.  The first phase is tunneling through a  multidimensional PES. The dynamics of this process, primarily adiabatic,  is governed by the collective fission inertia. Beyond the outer turning point, the system  propagates 
 in a classically 
allowed region before reaching scission, where it finally breaks into two 
fragments. The motion in the second phase has a dissipative  character. Consequently, the microscopic description of SF should involve  potential, inertial and  dissipative aspects \cite{(Swi72),(Ker74)}.

Although the tunneling phase could be described by instanton methods, 
which would account for some form of dissipation between collective and 
intrinsic degrees of freedom \cite{levit1980,negele1982,skalski2007},  numerous difficulties plague practical applications of the imaginary-time approach 
\cite{levit1980-a}. Consequently, most DFT-based calculations of tunneling probabilities 
are based  on the semiclassical WKB approximation and depend sensitively on the 
interplay between the static nuclear potential energy and the  collective inertia. 

The characterization of fission yields poses additional 
challenges. At scission, dissipation plays a crucial role, and 
would be best accounted for by time-dependent density functional theory (TDDFT). 
It is only very recently that  realistic time-dependent Hartree-Fock calculations of the fission process have become available \cite{simenel2014,scamps2015}. Albeit very promising, the current implementations of TDDFT treat  several 
important aspects of nuclear structure (center of mass, nuclear 
superfluidity) rather  crudely and cannot always properly describe collective correlations \cite{tanimura2015}. In addition, such calculations can only simulate single fission events: 
reconstructing the full mass distribution in TDDFT is beyond 
current computational capabilities, since it would involve large-scale 
Monte Carlo sampling of all possible fragmentations. Fortunately, such a 
sampling is easily doable within the classical Langevin dynamics \cite{abe1996}. 

The present work is an important milestone for our long-term project aiming at  providing  accurate description of SF within nuclear DFT. Recently, we demonstrated that the predicted SF pathways essentially depend on the assumptions behind the treatment of collective inertia and collective action \cite{(Sad13)}. In Ref.~\cite{(Sad14)}, we also showed that pairing dynamics can profoundly impact penetration probabilities by restoring symmetries spontaneously broken in the static approach.

In this Letter, we predict for the first time mass and 
charge distributions in SF within a unified theoretical 
framework. We employ state-of-the-art DFT to compute 
 adiabatic PESs and collective inertia in a 
multi-dimensional space of collective coordinates. This allows us to 
predict  the tunneling probabilities along the hypersurface of  outer turning points 
and  solve the Langevin equations to propagate the nucleus from the outer turning 
points to scission. The validity of such an approach is illustrated  in the 
benchmark case of $^{240}$Pu, where the experimental fission yields are well known  \cite{(Sal13),(Lai62),(Thi81)}.

{\it Theoretical framework} --- We calculate the SF half-life by following 
the formalism described in Ref.~\cite{(Sad13)}. In the WKB approximation, 
the half-life can be written as $T_{1/2}=\ln2/(nP)$ \cite{(Bar81),(Bar78)}, 
where $n$ is the number of assaults on the fission barrier per unit time 
(we adopt the standard value of $n=10^{20.38}s^{-1}$) and $P=1/(1+e^{2S})$ 
is the penetration probability expressed in terms of the action integral,
\begin{equation}
\label{action-integral}
S(L)
=
\int_{s_{\rm in}}^{s_{\rm out}}\sqrt{\frac{2\mathcal{M}_{\text{eff}}(s)}{\hbar^{2}}
\left( V(s) - E_0 \right)}\,ds ,
\end{equation}
calculated along the optimum fission path $L(s)$ connecting the inner and 
outer turning points $s_{\rm in}$ and $s_{\rm out}$ within a multi-dimensional 
collective space characterized by $N$ collective variables 
$\gras{q}= (q_{1},\dots,q_{N})$. The effective inertia $\mathcal{M}_{\text{eff}}(s)$ 
is obtained from the non-perturbative cranking inertia tensor 
$\mathcal{M}_{ij}$ \cite{(Bar81),(Bar78),(Bar05),(Sad13)}. The potential along the path is $V(s)$, and $E_0$ stands for 
the collective ground-state energy. 

At first, we compute the PES
$V(\gras{q})$ of the nucleus by solving the Hartree-Fock-Bogoliubov 
equations with constraints on $\gras{q}$. In order to stay consistent 
with our previous studies, we use the SkM* parametrization \cite{(Bar82)} 
of the Skyrme energy density and  a density-dependent mixed-type 
pairing term \cite{doba02}. The pairing strength is locally 
adjusted to reproduce odd-even mass differences \cite{(Sch14)}. Without missing crucial physics, we divide the collective space into two 3D regions to improve  the numerical efficiency of the calculation. In the region of elongations between the ground state and fission isomer, the most relevant 
degrees of freedom are the elongation, represented by the mass quadrupole moment $Q_{20}$; triaxiality, represented by the mass quadrupole moment $Q_{22}$; and
the coordinate $\lambda_2$ representing  dynamic pairing fluctuations~\cite{(Sad14)}. This results in a 3D space
$\gras{q}_{\rm I} \equiv (Q_{20},Q_{22},\lambda_2)$.
For elongations greater than that of the fission isomer, triaxiality plays a 
minor role but reflection-asymmetric degrees of freedom, 
represented by the mass octupole moment $Q_{30}$, become important; hence, in that  region, our collective space is $\gras{q}_{\rm II} \equiv (Q_{20},Q_{30},\lambda_2)$. In practical calculations, it is convenient to
introduce dimensionless coordinates $\{x_i\}$, where $x_i=q_i/\delta q_i$
and $\delta q_i$ are the scale parameters that are also used when
determining numerical derivatives of density matrices.
Here, we employ the values of $\delta q_i$ as in Refs.~\cite{(Sad14),(Bar11)}.

\begin{figure}[!htb]
\includegraphics[width=1\columnwidth]{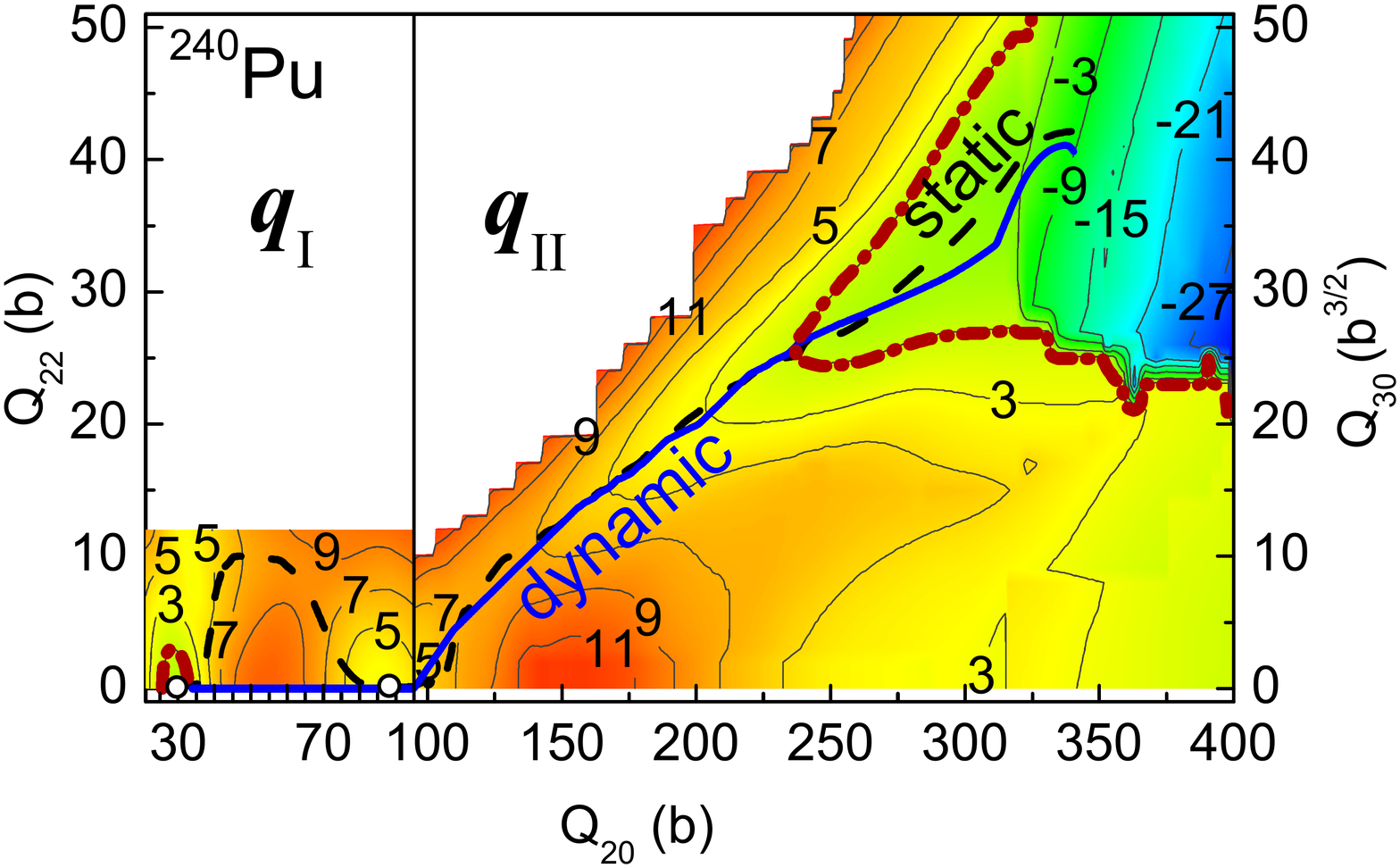}
\caption{\label{plot1}
(Color online) Projections of the static (dashed line) and dynamic (solid 
line) SF paths on the potential energy contours in the two considered regions
$\gras{q}_{\rm I}$ and $\gras{q}_{\rm II}$ of the collective space. The contours of inner and outer 
turning points are shown by dash-dotted lines. Ground-state and fission-isomer minima are marked by dots.}
\end{figure}

The potential  energy and inertia tensor are computed with the 
symmetry-unrestricted DFT solver HFODD (v2.49t)~\cite{(Sch12)}. 
The potential is corrected by subtracting the zero-point energy computed 
within the Gaussian overlap approximation \cite{(Sta89),(Bar07),(Sta13)}. 
The derivatives of the density matrix with respect to the collective 
coordinates, which are needed to compute the non-perturbative cranking inertia tensor, are calculated 
with the finite difference method~\cite{(Bar11)}. In Fig.~\ref{plot1}, 
we show the projections of the most probable fission path in the 
two-dimensional planes  $\{Q_{20}, Q_{22}\}$ and $\{Q_{20}, Q_{30}\}$. 
For all pairs of inner and outer turning points at energy $E_{0}$, the 
one-dimensional path $L(s)$ is calculated with the dynamic programming 
method~\cite{(Bar81)} by minimizing the action in the multidimensional space 
of $\{x_i\}$. In this way, we obtain a family of SF 
probabilities $P(s_{\rm out})$ that correspond to the hypersurface of outer turning  points $s_{\rm out}$.

For all the points $s_{\rm out}$, we then compute the time-dependent fission 
path to scission by solving the dissipative Langevin 
equations~\cite{abe1996,(Fro98)}:
\begin{align}
\label{langevin}
\frac{d p_i}{dt} &=
-\frac{p_j p_k}{2} \frac{\partial}{\partial x_i}(\mathcal{M}^{-1})_{jk}
- \frac{\partial V}{\partial x_i}\\ \nonumber
& - \eta_{ij}(\mathcal{M}^{-1})_{jk}p_k
+ g_{ij}\Gamma_j(t),\\ \nonumber
\frac{dx_i}{dt} &= (\mathcal{M}^{-1})_{ij}p_j,
\end{align}
where $p_i$ represents the momentum conjugate to $x_i$, 
$\eta_{ij}$ is the dissipation tensor, 
$g_{ij}\Gamma_j(t)$ is 
the random (Langevin) force with $\Gamma_j(t)$
being a time-dependent stochastic variable with a Gaussian distribution,
and $g_{ij}$  is the random-force strength tensor. The time-correlation property of the random force is 
assumed to follow the relation $\langle\Gamma_k(t)\Gamma_l(t')\rangle = 
2\delta_{kl}\delta(t-t')$. The strength of the random force is related 
to the dissipation coefficients through the fluctuation-dissipation 
theorem: $\sum_kg_{ik}g_{jk}=\eta_{ij}k_{\rm B}T$, where the temperature $T$ of 
the fissioning nucleus at any instant of its evolution is given by 
$k_{\rm B}T=\sqrt{E^*/a}$. Here, $a=A/10$\,MeV$^{-1}$ is the level density 
parameter and $E^*=V(s_{\rm out})-V(\gras{x})$ represents the excitation energy of the fissioning system system in the classically-allowed region;
for SF, $E^*$ increases as the system moves  toward scission beyond $s_{\rm out}$. Scission is 
defined here by the criterion that the number of particles in the neck 
between the two pre-fragments is less than 0.5. Each point on the 
scission hypersurface defines a split corresponding to two fission fragments. The mass and 
charge of the fragments are obtained from the calculated density distributions  \cite{(Sch14)}. Owing to the random force in the Langevin 
equations, repeating the calculation several times with the same initial 
condition at a given outer turning point $s_{\rm out}$  yields different trajectories: the 
charge and mass distribution are then simply obtained by counting the 
number of trajectories ending at a given fragmentation, weighing with 
$P(s_{\rm out})$, and normalizing the result to unit 
probability. Finally, to account for fluctuations of particle number 
in the neck at scission, Langevin yields are convoluted
with a Gaussian of width $\sigma$ \cite{younes2012-a}. 
Based on the expectation value of the total (proton) particle number in 
the neck region, we choose $\sigma=3$ (or 2) for $A$ (or $Z$). 

Compared to the Brownian shape-motion approach, which is applied to describe induced fission \cite{(Ran11),(Ran13),(Mol15)},
our model contains a number of attractive theoretical features: (i) it is based on a self-consistent theory utilizing realistic effective interactions both in particle-hole and particle-particle channels; (ii) the fission pathway is obtained by an explicit minimization of the collective action, i.e., the static assumption is not used; (iii) the inertial effects are considered both during tunneling and Langevin propagation; (iv) the full Langevin description of the nuclear shape dynamics is considered.

{\it Results} ---  The initial collective energy $E_0$ is a crucial 
quantity for determining the fission half-life.  To find $E_0$, we first calculate the most probable fission path of $^{240}$Pu by 
minimizing the action (\ref{action-integral}) in the 3D+3D space described above. An agreement with the experimental SF half-life \cite{(Sal13)} is 
achieved for $E_0=0.97$ MeV, which is indeed very close to a value of 1\,MeV assumed in our previous work
\cite{(Sad13),(Sad14)}.  In the following, 
 we adopt the value $E_0=0.97$ MeV to define the inner and outer turning 
points. As shown in Fig.~\ref{plot1} and discussed in detail in our previous work \cite{(Sad14)}, the least-action path  between the ground-state and  
fission isomer in $^{240}$Pu is axially-symmetric, and it dramatically differs from the static trajectory corresponding to the least energy path, which goes through triaxial shapes. In the region $\gras{q}_{\rm II}$, the dynamic path
is predicted to be very close to the static path. Note that in Fig.~\ref{plot1}, 
the part of the dynamic path outside the outer turning point is calculated by 
disregarding the random force in Eq.~(\ref{langevin}); this enforces deterministic trajectories (see also  below).

\begin{figure}[!htb]
\includegraphics[width=0.8\columnwidth]{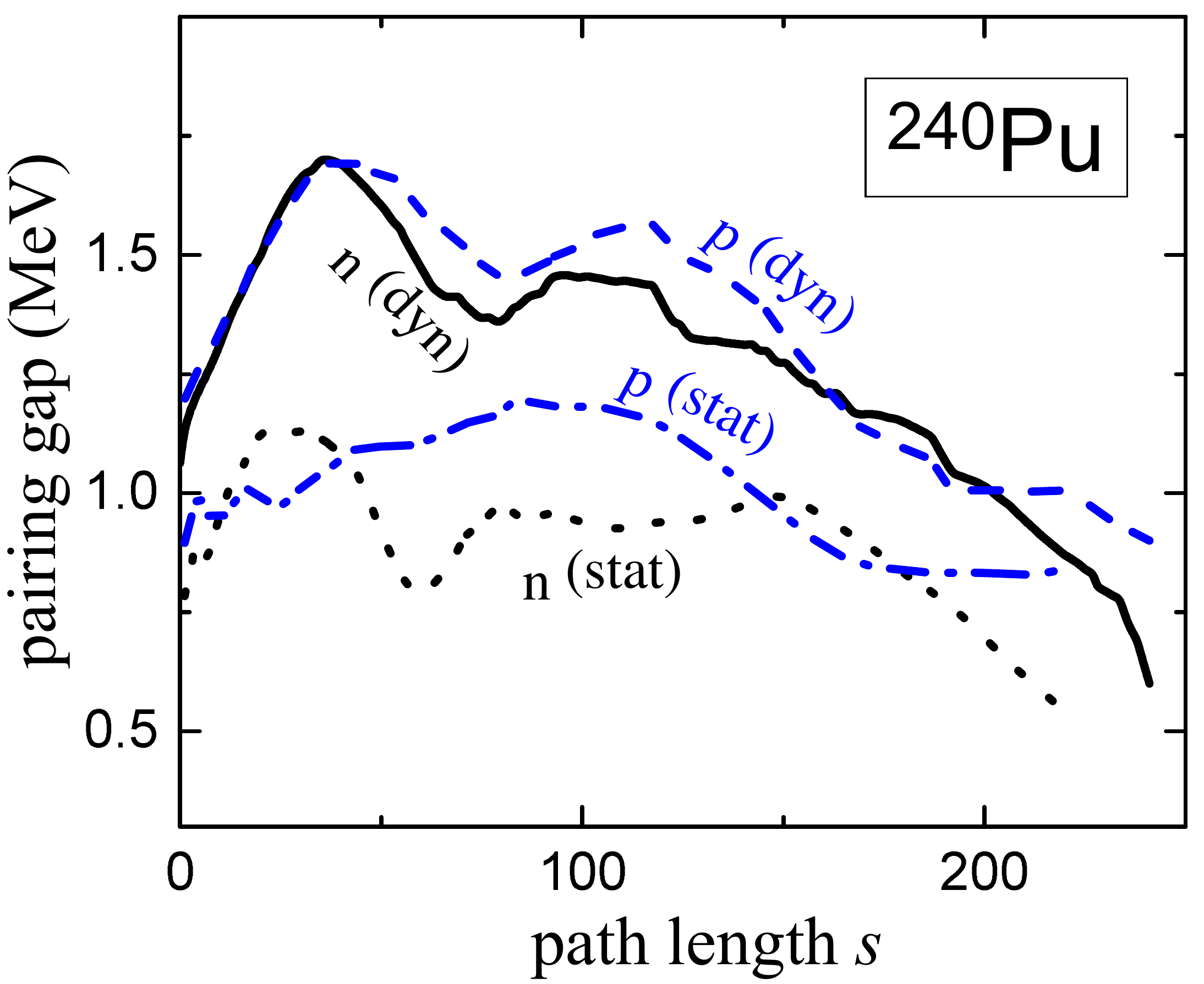}
\caption[C1]{\label{plot3}
(Color online) Variation of the pairing gap for neutrons and protons along 
the static and dynamic paths of Fig.~\ref{plot1}.}
\end{figure}

In general, we found that the impact of nuclear pairing on $S(L)$ becomes strongly reduced 
at large deformations and the pairing gaps attain the static values near 
the outer turning point. This is because the most probable fission pathways can be associated with shapes characterized by large symmetry breaking. This observation is demonstrated in Fig.~\ref{plot3} for the dynamic and static paths of Fig.~\ref{plot1} by showing the neutron and proton pairing gaps along the path length $s$. Subsequently, we restrict the dynamical space in the classically-allowed region to the surface defined by $\{Q_{20}, Q_{30}\}$. In the following, we calculate the fission paths on this surface for a 
collection of 900 outer turning points around the most probable $s_{\rm out}$.

\begin{figure}[!htb]
\includegraphics[width=1.0\columnwidth]{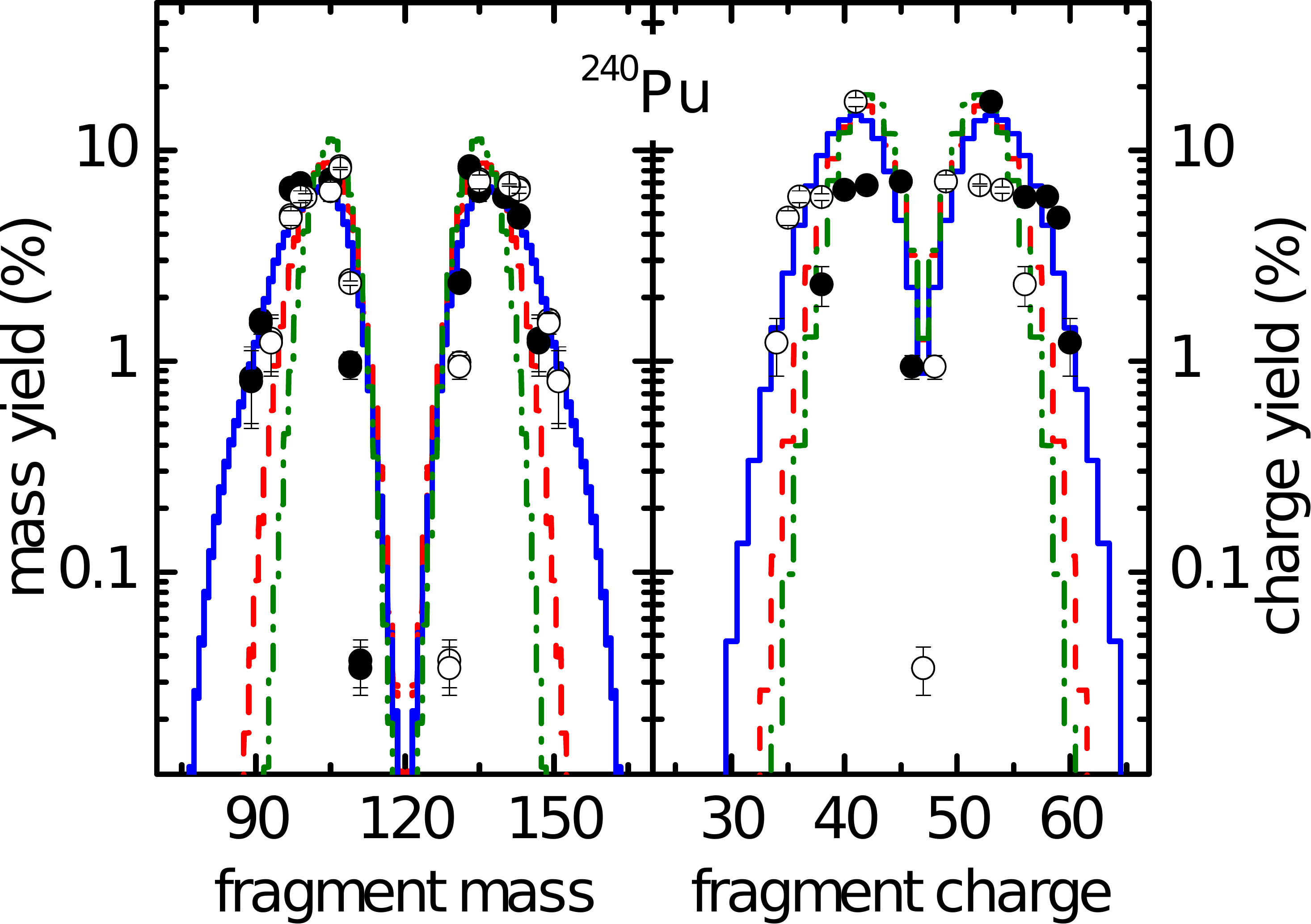}
\caption{\label{plot4}
(Color online) Mass (left) and charge (right) yield 
distributions for the SF of $^{240}$Pu. The experimental values~\cite{(Lai62),
(Thi81)} (mirror points) are shown by solid (open)  circles. Calculations 
with dissipative Langevin dynamics and full inertia (solid blue lines) are compared to results obtained with non-dissipative dynamics and full inertia  (red dashed lines) and  with dissipative 
dynamics and a diagonal unit inertia tensor  (green dashed-dotted lines).
}
\end{figure}

The Langevin propagation  is studied in three different scenarios. 
In the first variant, the mass and charge distributions of fission fragments are computed 
without invoking dissipation and fluctuation by setting $\eta_{ij} = 0$ (thus
 $g_{ij}=0$). 
Under such conditions, the Langevin equations resemble the deterministic 
Newtonian equations of motion with a one-to-one correspondence between 
outer turning points and scission points. By computing  900 trajectories to scission, we obtain mass and charge yield
distributions marked by the red dashed line in 
Fig.~\ref{plot4}. The most probable values of the fission yields are consistent with the data  but the distribution tails  are clearly off. In the second variant,  we incorporate a constant collective dissipation tensor 
$\eta_{ij}$ with reasonable values $\eta_{11}=50\hbar$, $\eta_{22}=40\hbar$ and 
$\eta_{12}=0$, but take a diagonal unit mass tensor and obtain the green dashed-dotted line. In this case, the fission dynamics is dominated by the static features of the PES: the maxima are reproduced, but not the width of the distributions. It is only by combining a constant dissipation tensor with the non-perturbative cranking inertia that we obtain the solid blue lines, which nicely agree with experiment over the whole range of mass/charge splits. The results shown in Fig.~\ref{plot4} correspond to 
100 different runs per each outer turning point, hence the distributions contain contribution from 90,000 trajectories. 

\begin{figure}[htb]
\includegraphics[width=1.0\columnwidth]{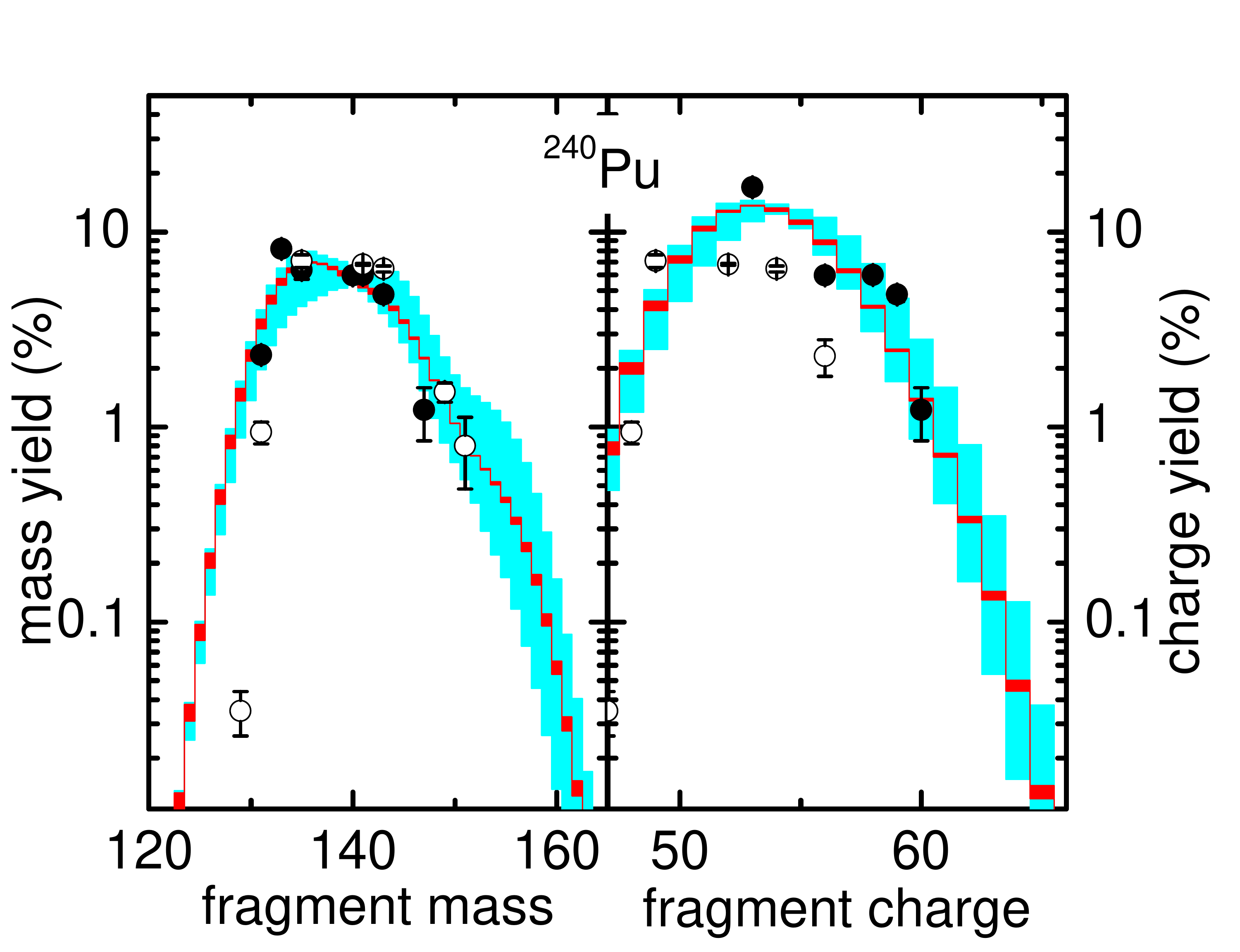}
\caption{\label{plot5}
(Color online) Mass (left) and charge (right) distributions 
of heavier SF yields of $^{240}$Pu. The symbols are the same as in Fig.~\ref{plot4}. 
The shaded regions are uncertainties in the distributions due to
variations in  $E_{0}$ (narrow red band) and dissipation tensor (wider cyan band).}
\end{figure}

To illustrate the sensitivity of yield distributions to
the initial 
collective energy $E_{0}$, the narrow red band
in Fig.~\ref{plot5}  shows the distribution
uncertainty  when taking a sample of 11 different values of $E_{0}$ 
within the range $0.7 \leq E_{0} \leq 1.2$\,MeV.
While such a  variation in $E_0$ 
changes the  SF half-life by over two orders of magnitude, 
its  impact on fission yield 
distributions is minimal. The wider cyan band shows the spread in 
predicted distributions when sampling the dissipation tensor  in the range of
$0\le \eta_{12}\le 30\hbar$  and 
$(\eta_{11},\eta_{22})\in [ 30\hbar,400\hbar]$ with the constraint 
$1\le \eta_{11}/\eta_{22} \le 1.25$. Note that we consider a very 
broad range of variations in order to account for the 
uncertainties in the theoretical determination the dissipation tensor. It is very encouraging to see that the predicted yield distributions vary relatively  
little, even for   nonphysically large values  of $\eta_{ij}$. 

{\it Conclusions} --- In this letter, we propose a microscopic approach rooted in nuclear DFT 
to calculate mass and charge distributions of SF yields. The
SF penetrabilities,  obtained by minimizing the collective action in large multidimensional PESs with realistic collective inertia,  
are used as inputs to solve the time-dependent dissipative
Langevin equations. By combining many trajectories connecting the hypersurface of outer turning points with the scission hypersurface, we predict SF yield distributions. The results of our pilot calculations for $^{240}$Pu   are in excellent agreement with experiment and remain quite stable 
under large variations of input parameters, such as the collective energy $E_0$ or the dissipation tensor. This is an important outcome, as SF yield distributions
are important observables for benchmarking theoretical models of SF \cite{(Ber15)}. This finding is reminiscent of the analysis of Ref.~\cite{(Ran11)} for low-energy neutron- and gamma-induced fission, which found that the yield distributions predicted in the Brownian-motion approach  are insensitive to large variations of dissipation tensor. On the other hand, according to our analysis, the collective inertia tensor impacts both tunneling and  the Langevin dynamics. 

The results of our study confirm that the PESs is the most important ingredient when it comes to the maxima of  yield distributions. This is consistent with the previous DFT studies of most probable SF splits \cite{Sta09,(War12a),mcdonnell2013,(McD14),scamps2015-a},
which indicate that the topology of the PES in the pre-scission region is the crucial factor. On the other hand, both dissipative collective dynamics and collective inertia are essential when it comes to the shape of the yield distributions. The fact that the predictions are fairly robust with respect to the details of dissipative aspects of the model is most encouraging.

\begin{acknowledgments}
Discussions with A. Baran, J. Dobaczewski, J. A. Sheikh and S. Pal are
gratefully acknowledged. This work was supported by the U.S. Department
of Energy, Office of Science, Office of Nuclear Physics under Award
Numbers 
DOE-DE-NA0002574 (the Stewardship Science Academic Alliances
program) and   DE-SC0008511 (NUCLEI SciDAC-3 collaboration). 
Part of this research
was performed under the auspices of the U.S.\ Department of Energy by
Lawrence Livermore National Laboratory under Contract DE-AC52-07NA27344.  
Computational resources were provided through
an INCITE award ``Computational Nuclear Structure" by the
National Center for Computational Sciences (NCCS), and by
the National Institute for Computational Sciences (NICS). Computing resources were also provided through an award
by the Livermore Computing Resource Center at Lawrence Livermore National
Laboratory.
\end{acknowledgments}

\bibliographystyle{apsrev4-1}
\bibliography{ref,zotero_output,books}

\end{document}